\documentclass[10pt]{article}

\usepackage[margin=1in]{geometry}
\usepackage{amsmath}
\usepackage{amssymb}
\usepackage{amsthm}
\usepackage{graphicx}
\usepackage{booktabs}
\usepackage{rotating}
\usepackage{epstopdf}% To incorporate .eps illustrations using PDFLaTeX, etc.
\usepackage{enumitem}
\usepackage[caption=false]{subfig}% Support for small, `sub' figures and tables
\usepackage{xcolor}
\usepackage{comment}
\usepackage{makecell}
\usepackage{tabularx}
\usepackage{placeins}
\usepackage{afterpage}

\usepackage[numbers,sort&compress]{natbib}
\usepackage[hidelinks]{hyperref}

\theoremstyle{plain}% Theorem-like structures provided by amsthm.sty

\theoremstyle{definition}

\theoremstyle{remark}

\title{BrainNormalizer: Anatomy-Informed Pseudo-Healthy Brain Reconstruction from Tumor MRI via Edge-Guided ControlNet}

\author{
Min Gu Kwak\textsuperscript{1,2}\thanks{Corresponding author: Jing Li. Email: jli3175@gatech.edu}
\and Yeonju Lee\textsuperscript{2}
\and Hairong Wang\textsuperscript{2,3}
\and Kristin R. Swanson\textsuperscript{4,5}
\and Jing Li\textsuperscript{2,*}
}

\date{}

\begin{document}

\maketitle

\begin{center}
\small
\textsuperscript{1}Department of Health and Rehabilitation Sciences, University of Pittsburgh, Pittsburgh, PA, USA\\
\textsuperscript{2}H. Milton Stewart School of Industrial and Systems Engineering, Georgia Institute of Technology, Atlanta, GA, USA\\
\textsuperscript{3}Operations Research and Industrial Engineering, Cockrell School of Engineering, University of Texas at Austin, Austin, TX, USA\\
\textsuperscript{4}MOSAIC Center, Cedars-Sinai Medical Center, Los Angeles, CA, USA\\
\textsuperscript{5}Computational Biomedicine and Pathology/Lab Medicine, Department of Neurosurgery, Mayo Clinic Arizona, Phoenix, AZ, USA
\end{center}

\begin{abstract}
Brain tumors induce complex structural deformations that obscure the patient’s original neuroanatomy, making it difficult to distinguish tumor-induced changes from inherent anatomical variability. Reconstructing a subject-specific pseudo-healthy brain can provide a critical reference for such analysis, but this task is inherently counterfactual, as paired pre-tumor scans and explicit healthy guidance are unavailable.
We propose BrainNormalizer, a diffusion-based framework for subject-specific pseudo-healthy brain MRI reconstruction that enables anatomy-informed reconstruction without requiring paired data or explicit healthy references. The framework learns anatomical priors and edge-based structural conditioning through a two-stage training strategy consisting of inpainting-based diffusion fine-tuning and ControlNet-based edge conditioning. At inference, counterfactual pseudo-healthy reconstruction is achieved through a deliberate misalignment strategy, where tumorous inputs are paired with non-tumorous prompts and mirrored contralateral edge maps. This allows subject-specific anatomical guidance to be constructed from the patient’s own anatomy, enabling anatomically consistent pseudo-healthy reconstruction that preserves individual structural characteristics.
Experiments on the BraTS2020 dataset demonstrate that BrainNormalizer achieves improved distributional realism, symmetry-based structural consistency, and reduced false positive detection compared to existing methods. These results indicate that the proposed framework provides a principled approach for subject-specific counterfactual reconstruction and supports downstream analysis of tumor-induced deformation.
\end{abstract}

\noindent\textbf{Keywords:} Diffusion model; ControlNet; brain MRI; pseudo-healthy reconstruction; medical image synthesis

\section{Introduction}

\label{sec:intro}
Brain tumors are among the most life-threatening and clinically significant neurological diseases that affect individuals across all age groups worldwide \cite{schaff2023glioblastoma,stupp2005radiotherapy}. Their aggressive biological behavior and structural heterogeneity lead to severe neurological impairments and poor clinical outcomes, making them a major cause of morbidity and mortality \cite{ostrom2022cbtrus}. This inherent complexity poses significant challenges for both clinical interpretation and computational analysis \cite{ce2023artificial}.
Furthermore, it leads to profound alterations in the structural and functional organization of the brain, thereby complicating accurate diagnosis and treatment planning \cite{falco2024functional, singh2021radiomics}. Magnetic Resonance Imaging (MRI) has become the primary modality for brain tumor assessment due to its superior soft-tissue contrast, high spatial resolution, and radiation-free imaging capability compared with Computed Tomography (CT) or other modalities \cite{mabray2015modern, fink2015multimodality}. MRI enables detailed visualization of tumor-induced deformations and subtle structural variations in surrounding healthy tissues, which makes it indispensable for diagnosis, treatment planning, and surgical guidance \cite{bauer2013survey}. 

Accurate interpretation of brain MRIs is essential for clinical tasks such as tumor delineation, treatment planning, and surgical decision-making \cite{lakhani2023current}. However, due to substantial inter-individual anatomical variability, distinguishing tumor-induced changes from a patient’s inherent neuroanatomy remains challenging \cite{seghier2018interpreting}. Furthermore, many automated MRI analysis methods are primarily designed for healthy tissue, and pathological regions are often simply masked or replaced during preprocessing, which limits reliable analysis in tumor-affected areas \cite{durrer2024denoising}. A patient-specific healthy reference would thus provide a critical basis for comparison, enabling more precise identification of tumor-induced deformation and improving the reliability of both clinical interpretation and computational analysis \cite{falco2024functional}. This issue is not limited to brain tumors and also arises in broader medical imaging settings, where pre-disease states are often unavailable. 

Such a reference would enable subject-specific quantification of structural deformations and facilitate longitudinal studies that are otherwise infeasible \cite{baheti2021brain}. However, obtaining it from patients is inherently difficult. Patients rarely undergo MRI scans before symptom onset, and the high cost of MRI with limited insurance coverage makes routine screening impractical \cite{schmidt2022effects}. As paired pre- and post-tumor scans are nearly impossible to obtain, there is strong motivation to computationally synthesize a pseudo-healthy MRI reflecting how the patient's brain would appear in the absence of tumor-induced alterations.

Generating high-quality images is inherently challenging because visual data contain complex contextual and structural relationships that must remain semantically and spatially consistent. Traditional generative models such as GANs \cite{goodfellow2020generative} and autoencoder-based frameworks \cite{wang2018unregularized} have shown potential in image synthesis, but often suffer from instability and structural fidelity issues \cite{cobbinah2025diversity}. These limitations are especially critical in medical imaging, where anatomical precision is essential for clinical interpretation \cite{yu2020medical}.

Diffusion models have since emerged as a more stable and expressive alternative \cite{ho2020denoising}, with latent diffusion model (LDM)-based frameworks such as Stable Diffusion (SD) enabling high-quality, semantically coherent synthesis through efficient latent-space denoising \cite{rombach2022high}. They have been applied to a wide range of tasks, including image synthesis, translation, segmentation, and anomaly detection \cite{wolleb2022diffusion,khader2023denoising,ozbey2023unsupervised}.
Recent studies have incorporated anatomical priors into diffusion-based medical image generation using structure-aware control frameworks such as ControlNet \cite{konz2024anatomically, xie2025meddiff, zhang2023adding}, enhancing anatomical fidelity through spatial conditioning with edge maps, segmentation priors, or shape constraints. However, counterfactual pseudo-healthy reconstruction from tumorous MRI remains underexplored, and existing approaches lack explicit mechanisms to incorporate subject-specific anatomical priors when inferring a plausible healthy state that is absent from the available data.

In this study, we present BrainNormalizer, a ControlNet-based diffusion framework designed to reconstruct subject-specific pseudo-healthy brain MRIs from tumorous scans while maintaining anatomical fidelity. Our model generates pseudo-healthy brain structures through a two-stage training scheme. In the first stage, an inpainting-based fine-tuning scheme adapts the pretrained diffusion model from natural images to the MRI domain, enabling anatomically coherent reconstruction of pseudo-healthy and tumorous tissues. In the second stage, the ControlNet branch is trained with structural cues extracted from edge maps, which highlight key anatomical boundaries such as cortical folds, ventricular contours, and major tissue interfaces. Conditioning the diffusion process on these boundary cues constrains generation to follow stable neuroanatomical geometry. This helps yield pseudo-healthy reconstructions that are more anatomically plausible and structurally consistent than those produced by unconstrained diffusion alone. During inference, counterfactual pseudo-healthy brain generation is achieved by deliberately misaligning tumorous MRIs with non-tumorous guidance through non-tumorous prompts and mirrored contralateral edge maps. This design leverages contralateral structural cues to generate anatomically consistent and clinically interpretable reconstructions without requiring paired scans from pre-tumor and tumorous states. The main contributions of this study are summarized as follows:
\begin{itemize}
    \item 
    {\textbf{Subject-specific anatomy-aware reconstruction for counterfactual pseudo-healthy brain MRI generation.}
    We introduce a framework for pseudo-healthy brain MRI generation that explicitly targets subject-specific and anatomically consistent reconstruction under tumor-induced structural deformation. Pseudo-healthy reconstruction is inherently a counterfactual problem, where neither paired pre-tumor scans, historical healthy references, nor explicit healthy guidance are available, while requiring consistency with patient-specific anatomy under substantial inter-subject variability. The proposed approach focuses on preserving individual neuroanatomical characteristics, enabling reconstructions that reflect patient-specific structure rather than generic healthy appearance.}
    \item
    {\textbf{Deliberate misalignment for supervision-free anatomical guidance construction.}}
    {
    We address the lack of both paired data and explicit healthy guidance by introducing a deliberate misalignment strategy at inference time. Rather than relying on ground-truth supervision, the model is driven by guidance signals constructed from non-tumorous prompts and contralateral structural cues, enabling subject-consistent pseudo-healthy reconstruction. This mechanism allows the model to integrate learned normal anatomical priors with patient-specific structure, even when no direct healthy reference is available.
    }
    \item
    \textbf{Clinical and research implications for brain tumor assessment and treatment planning.}
    From a clinical perspective, BrainNormalizer offers anatomically consistent reference images that approximate how the subject’s brain might appear without tumor-induced changes, thereby serving as a pseudo-healthy structural baseline. This facilitates more accurate tumor delineation, supports preoperative planning that preserves critical regions, and enables postoperative and longitudinal evaluations. Furthermore, by generating anatomically reliable pairs of pseudo-healthy and tumorous MRI, BrainNormalizer provides valuable resources for downstream studies, including longitudinal analysis, counterfactual modeling, and quantitative assessment of tumor-induced deformation.
\end{itemize}
%%%%%%%%%%%%%%%%%%%%%%%%%%%%%%%%%%%%%%%%%%
\section{Related Works}
\label{sec:literature}
\subsection{Diffusion models for medical imaging tasks}
\label{sec:literature-diffusion}
Diffusion models have been applied for medical image analysis across diverse tasks and imaging modalities \cite{hung2023med, zhang2025diffuseg, khader2023denoising, zhang2023adding, wu2024medsegdiff, wolleb2022diffusion, zhang2024diffboost}. Hung et al. (2023) proposed Med-cDiff, a conditional diffusion probabilistic model for medical image super-resolution, denoising, and inpainting \cite{hung2023med}. Zhang et al. (2025) proposed DiffuSeg, a conditional diffusion model that synthesizes medical images from label maps and unlabeled target-domain data to mitigate annotation costs and improve segmentation performance across domains \cite{zhang2025diffuseg}. Khader et al. (2023) proposed a diffusion probabilistic framework for synthesizing realistic and anatomically consistent medical images from MRI and CT data \cite{khader2023denoising}. Beyond image synthesis, diffusion-generated data have also been utilized to enhance segmentation, leveraging their probabilistic and generative nature to enable step-wise refinement and anatomically consistent predictions across modalities \cite{wu2024medsegdiff}. Furthermore, Wolleb et al. (2022) introduced a weakly supervised anomaly detection framework based on DDIM, using deterministic noising–denoising with classifier guidance to translate between tumorous and pseudo-healthy images and produce detailed anomaly maps without complex training \cite{wolleb2022diffusion}. Furthermore, with the advent of large language models (LLMs), text-to-image diffusion models have been introduced \cite{zhang2024diffboost}, allowing intuitive and semantic control of generated images through text prompts.

\subsection{ControlNet}
\label{literature-controlnet}
Researchers have investigated to control diffusion models on various forms of external information, such as class labels, textual prompts, or structural maps, to achieve controlled image generation. Such conditional diffusion frameworks enable both semantic and spatial guidance. Among these approaches, ControlNet has emerged as a powerful framework that injects structural conditions into large pretrained diffusion models while preserving their generative capabilities \cite{zhang2023adding}.

ControlNet has inspired a number of extensions in medical imaging, where preserving anatomical consistency is essential. Konz et al. (2024) proposed a ControlNet-based diffusion model for breast MRI generation, where multi-class anatomical segmentation masks served as spatial guidance during sampling \cite{konz2024anatomically}. Wang et al. (2025) proposed 3D MedDiffusion, demonstrating the strong potential of diffusion-based frameworks for CT and MRI tasks, where control is applied through depth and spatial structure conditions \cite{wang20253d}. {Zhao et al. (2025) introduced a locally controllable diffusion framework for image generation that enables region-specific conditioning through regional discriminate loss and feature mask constraints with ControlNet \cite{zhao2025local}.} Furthermore, researchers applied a ControlNet framework for lung MRI-to-CT translation \cite{rue2025misalignment}, using multi-channel conditioning that combined edge maps and intensity-normalized representations to improve alignment of structural boundaries. Recent study  incorporated ControlNet to improve medical image segmentation, enforcing image–mask consistency via a noise consistency loss \cite{qiu2025noise}. 

\subsection{Pseudo-healthy medical image  generation}
\label{literature-tumor}
{Pseudo-healthy medical image generation has been increasingly investigated to enable analysis under a healthy anatomical assumption. However, this formulation is inherently ill-posed due to the absence of paired ground-truth healthy images for the same subject, often leading to anatomically inconsistent reconstructions.}

{To address this challenge, various strategies have been proposed to improve reconstruction quality. A weakly supervised anomaly detection framework based on DDIM employs classifier guidance to translate pathological images into pseudo-healthy representations \cite{wolleb2022diffusion}. Diffusion-based inpainting methods reconstruct tumor-affected regions conditioned on masked inputs \cite{durrer2024denoising}. Extensions such as hierarchical diffusion improve multi-scale consistency \cite{kwark2025hierarchical}, while wavelet-based diffusion enhances structural detail preservation in the frequency domain \cite{ferreira2024brain}. Additional approaches introduce guidance or structured representations to improve anatomical plausibility. For instance, researchers guided the denoising process through anomaly-aware objectives to preserve healthy regions while suppressing pathological structures \cite{bercea2024diffusion}. Disentangled generative models separate normal and pathological components in latent space, enabling reconstruction by removing disease-related features \cite{li2025disentangled}. Hybrid frameworks further approximate supervision by generating pseudo-paired data via adversarial models, which are then used to train conditional diffusion models for reconstruction \cite{wang2024two}.
}

\subsection{Limitations of Existing Approaches}
\label{literature-lim}
{Despite recent advances in diffusion-based medical image synthesis and the introduction of controllable frameworks such as ControlNet, existing pseudo-healthy reconstruction methods predominantly rely on learned mappings or population-level priors to generate anatomically plausible healthy images. While visually realistic, these approaches do not explicitly ensure consistency with the patient’s own anatomy under substantial inter-subject variability.} {In particular, these methods do not address how to incorporate subject-specific anatomical structure into reconstruction in the absence of paired data or explicit healthy guidance. As a result, the generated outputs are largely driven by population-level statistics rather than being grounded in the individual patient’s anatomy, often leading to anatomically inconsistent or overly generic reconstructions.}

{To address this limitation, we propose BrainNormalizer, which enables subject-specific pseudo-healthy reconstruction by leveraging subject-specific anatomical structure derived from the patient’s own anatomy.}
%%%%%%%%%%%%%%%%%%%%%%%%%%%%%%%%%%%%%%%%%%
\section{Preliminaries}

\label{sec:prelim}
\begin{figure}[!htbp]
    \centering
    \includegraphics[width=0.58\linewidth,height=0.45\textheight,keepaspectratio]{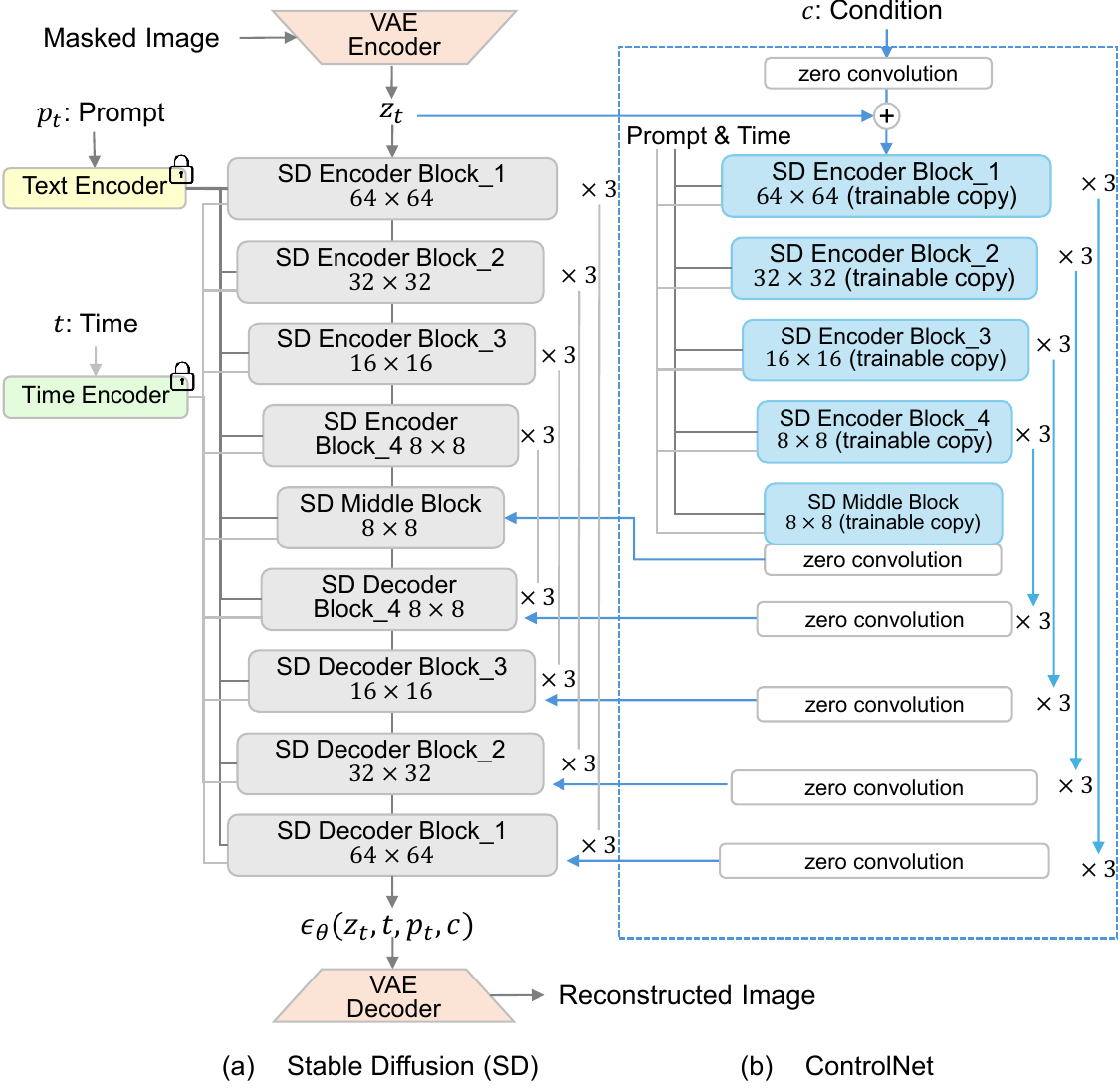}
    \caption{
    Overall architecture of the ControlNet. 
    (a) Stable Diffusion (SD) backbone that performs latent-space denoising conditioned on text and time embeddings. 
    (b) ControlNet branch that processes structural edge maps and injects boundary-aware features into the frozen SD decoder via zero-convolution layers. \textbf{Abbreviations:} VAE, Variational Autoencoder; SD, Stable Diffusion.
    }
    \label{figure:architecture}
\end{figure}

{ControlNet is a large pretrained text-to-image diffusion models that are controlled by conditioning signals such as edge maps or segmentation maps with end-to-end learning \cite{zhang2023adding}. Figure \ref{figure:architecture} presents the overall ControlNet architecture, which consists of two branches: a SD backbone and a parallel trainable ControlNet branch for conditioning. Figure \ref{figure:architecture}(a) illustrates the architecture of SD, which performs the denoising process in a latent space. An input image is first encoded into a compact latent representation \( z_t \) through a variational autoencoder (VAE). The text prompt $p_t$ is converted into a text embedding using a CLIP text encoder and, together with the diffusion timestep $t$, conditions a U-Net that predicts the noise added to the latent variable $\epsilon_{\theta}(z_t, t, p_t)$. The denoised latent by this predicted noise is then mapped back to the image space by the VAE decoder. As shown in Figure \ref{figure:architecture}(b), the ControlNet branch duplicates the encoder and middle blocks of the diffusion U-Net and extracts multi-scale control features from the conditioning signals. These features are passed through zero-initialized convolution layers and injected into the corresponding decoder stages of the backbone shown in Figure \ref{figure:architecture}(a) via residual connections. As a result, the noise prediction becomes conditioned on both the text prompt and the conditioning signal, which can be expressed as \( \epsilon_\theta(z_t, t, p_t, c) \). The strength of how strongly the model follows the conditioning signal can be modulated by a control parameter.}

%%%%%%%%%%%%%%%%%%%%%%%%%%%%%%%%%%%%%%%%%%
\FloatBarrier
\section{Materials and Methods}
\label{sec:method}
\subsection{Overview}
\label{sec:method-overview}
{We propose BrainNormalizer, a framework for reconstructing subject-specific pseudo-healthy brain MRIs by generating plausible counterfactual anatomy in tumor-affected regions without relying on paired pre-tumor scans or explicit healthy references. This aims to approximate how the brain might appear without tumor-induced deformation, while preserving subject-specific anatomical structure.}
The overall architecture of BrainNormalizer follows the ControlNet framework as illustrated in Figure \ref{figure:architecture}.

BrainNormalizer is trained to reconstruct subject-specific pseudo-healthy MRIs from tumorous ones by imposing anatomical control guided by edge-based structural cues.
In the first stage, the pretrained Stable Diffusion v1.5, originally trained on large-scale natural image datasets, is fine-tuned for domain adaptation to brain MRIs through inpainting-based reconstruction, as shown in Figure~\ref{figure:step1}. 
Subsequently, a ControlNet is trained to condition the diffusion process on structural guidance derived from Canny edge maps~\cite{canny2009computational} with the SD weights from the first stage kept frozen, as illustrated in Figure~\ref{figure:step2}. 
Finally, BrainNormalizer reconstructs a pseudo-healthy brain MRI from a tumorous MRI by leveraging a misaligned healthy prompt and a mirrored edge map, which jointly guide the model toward anatomically consistent, pseudo-healthy MRI reconstruction.
\subsection{Stable Diffusion Fine-tuning}
\label{sec:method-sdfinetune}
\begin{figure}[!htbp]
    \centering
    \includegraphics[width=0.8\linewidth]{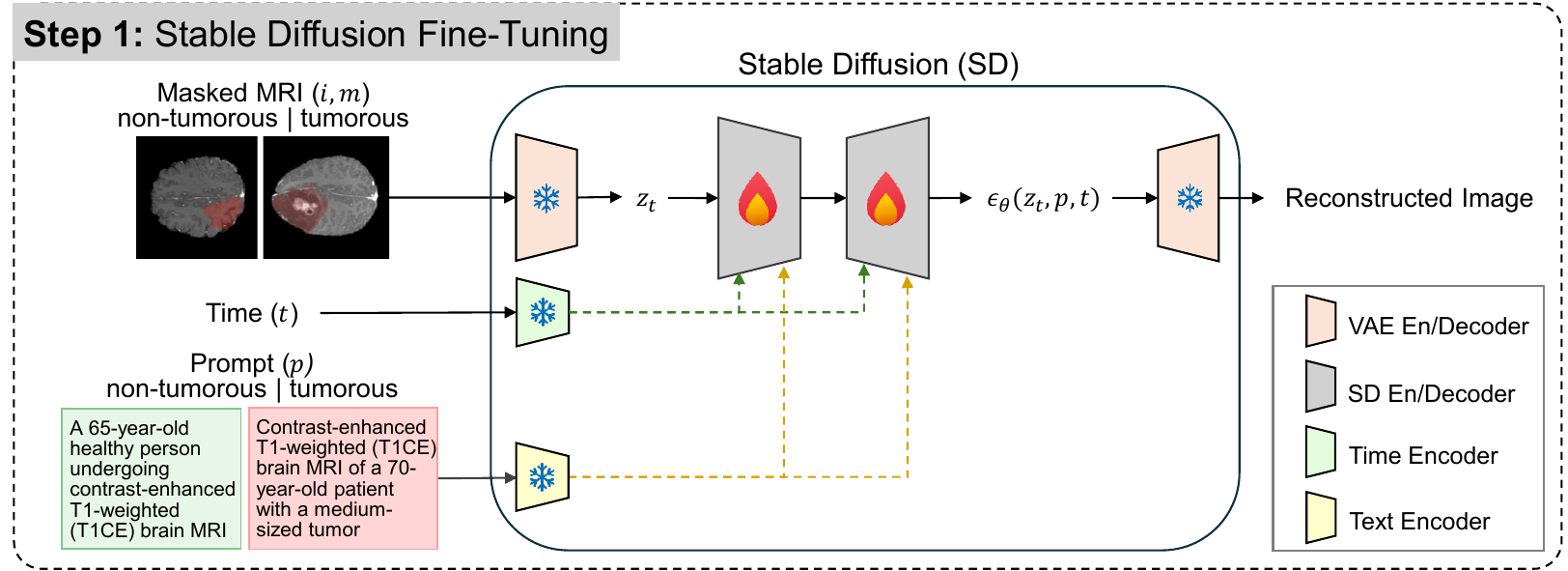}
    \caption{
    Step 1: Stable Diffusion (SD) fine-tuning via inpainting. 
Masked inputs $(i, m)$ with prompt $p$ and timestep $t$ are mapped to latent $z_t$ and denoised via $\epsilon_\theta(z_t, p, t)$. 
The model reconstructs masked regions consistent with the input condition (tumorous or non-tumorous).
} 
    \label{figure:step1}
\end{figure}
As illustrated in Figure~\ref{figure:step1}, the first stage of BrainNormalizer fine-tunes the pretrained SD to adapt it to the medical MRI domain through inpainting-based reconstruction.
The SD backbone introduced in Section~\ref{sec:method-overview} is used here with its VAE, denoising U-Net, and cross-attention conditioning modules.
During this stage, we freeze the pretrained VAE, CLIP text encoder, and noise scheduler, and fine-tune only the U-Net parameters to learn domain-specific MRI priors.

Both non-tumorous and tumorous cases are provided as independent input pairs \((i, m, p, t)\),  
where \(i\) denotes the input 2D MRI slice, \(m\) the binary mask defining the inpainting region, \(p\) the text prompt, and \(t\) the diffusion timestep.  
For non-tumorous cases, the inputs are represented as \((i_n, m_n, p_n, t)\), where \(i_n\) is a non-tumorous MRI slice, \(m_n\) is the corresponding mask, and \(p_n\) is a prompt describing a normal brain structure.  
For tumorous cases, the inputs are represented as \((i_d, m_d, p_d, t)\), where \(i_d\) is a tumorous MRI slice, \(m_d\) is a binary tumor mask, and \(p_d\) is a prompt describing the presence and size of the tumor.  
Each input is constructed to reflect its clinical context, ensuring that the model learns to reconstruct anatomically plausible structures consistent with the semantic and spatial cues provided.

The binary mask \(m\) specifies the region of the image to be inpainted.  
For tumorous cases, the mask was derived from the ground-truth binary tumor mask and dilated by five pixels to ensure full coverage of the tumor boundaries.  
For non-tumorous cases, since no tumor masks are available, a mask was randomly borrowed from a tumorous slice with the same slice index to maintain spatial consistency and enhance robustness to mask variations.  
The masked image is obtained by applying \(m\) to \(i\). This serves as the model input, where the masked region indicates the area to be reconstructed while the unmasked regions provide anatomical context for inpainting. During training, the loss is computed only within the masked region, ensuring that updates are focused on reconstructing the missing content while preserving the surrounding context.

The text prompt \(p\) provides semantic conditioning for each case and was generated from fixed template sets corresponding to the non-tumorous and tumorous cases. For both cases, a prompt was randomly selected from three predefined templates. {This random sampling across prompt templates acts as text-side augmentation, encouraging the model to learn phrasing-invariant representations and improving robustness to linguistic variations \cite{saharia2022photorealistic}.} For both healthy and tumorous MRIs, text prompts were generated from predefined template sets with parameterized attributes.

For \textbf{non-tumorous MRIs}, prompts describe normal brain anatomy without pathological findings, parameterized by imaging modality and subject age:
{\setlength{\baselineskip}{0.8\baselineskip}
\begin{itemize}
    \setlength{\itemsep}{0pt}
    \item \{modality\} of a \{age\_desc\} healthy individual
    \item A \{age\_desc\} healthy person undergoing \{modality\}
    \item \{modality\} image of a healthy brain (age: \{age\_desc\})
\end{itemize}
}

For \textbf{tumorous MRIs}, prompts additionally encode tumor characteristics, parameterized by imaging modality, subject age, and tumor size:
{\setlength{\baselineskip}{0.8\baselineskip}
\begin{itemize}
    \setlength{\itemsep}{0pt}
    \item \{modality\} of a \{age\_desc\} patient with a \{size\_desc\} tumor
    \item A \{age\_desc\} patient’s \{modality\} scan showing a \{size\_desc\} tumor
    \item \{modality\} image showing a \{size\_desc\} brain tumor in a \{age\_desc\} patient
\end{itemize}
}

Tumor size categories were defined based on pixel counts: small ($<1{,}400$), mild ($1{,}400$--$1{,}800$), medium ($1{,}800$--$2{,}100$), moderate ($2{,}100$--$2{,}500$), and large ($>2{,}500$). When age information was unavailable, a placeholder (\textit{unknown age}) was used.

The diffusion timestep $t$ represents the noise level in the forward diffusion process, determining the degree of corruption applied to the latent representation before denoising-based reconstruction.

As depicted in Figure~\ref{figure:step1}, each masked input image is encoded into a latent representation \(z_0\) using the pretrained VAE encoder. Gaussian noise $\epsilon \sim \mathcal{N}(0, I)$ is added at a randomly sampled timestep $t$, producing a noisy latent $z_t$ via the diffusion forward process $z_t = \sqrt{\bar{\alpha}_t}\, z_0 + \sqrt{1 - \bar{\alpha}_t}\,\epsilon$, where $\bar{\alpha}_t$ is the cumulative product of the noise schedule coefficients. The text prompt \(p\) is transformed into text embeddings via the pretrained CLIP text encoder, and the timestep \(t\) is encoded using sinusoidal positional encoding.  
Both embeddings are injected into the U-Net through cross-attention and residual connections to provide semantic and temporal conditioning.  
The U-Net then predicts the added noise term as \(\epsilon_\theta(z_t, p, t)\) based on these encoded representations.
The loss function of SD fine-tuning minimizes the discrepancy between the true noise \(\epsilon\) and the predicted noise \(\epsilon_\theta(z_t, p, t)\) only inside the masked region, formulated as:
\[
\mathcal{L}_{\text{SD}} = 
\mathbb{E}_{z_t, \epsilon, t, m}
\left[
\|\, m \odot (\epsilon - \epsilon_\theta(z_t, p, t)) \,\|_2^2
\right],
\]
where \(\odot\) represents element-wise multiplication that restricts the loss computation to the masked area. 
After denoising, the latent is decoded by the VAE to reconstruct the full MRI image, with the masked region restored and the surrounding anatomy preserved.  

By inpainting-based fine-tuning, the model learns to reconstruct both non-tumorous and tumorous cases in accordance with their semantic and anatomical contexts, leveraging text conditioning and pretraining.

\subsection{ControlNet Training}
\label{sec:method-controlnet}

After SD is fine-tuned through inpainting to capture general anatomical priors, the ControlNet is subsequently trained to provide subject-specific anatomical guidance using edge maps extracted from the input MRIs. These edge maps encode the major structural contours of the brain and help preserve the subject’s underlying anatomical layout.

\begin{figure}[!htbp]
    \centering
    \includegraphics[width=0.80\linewidth]{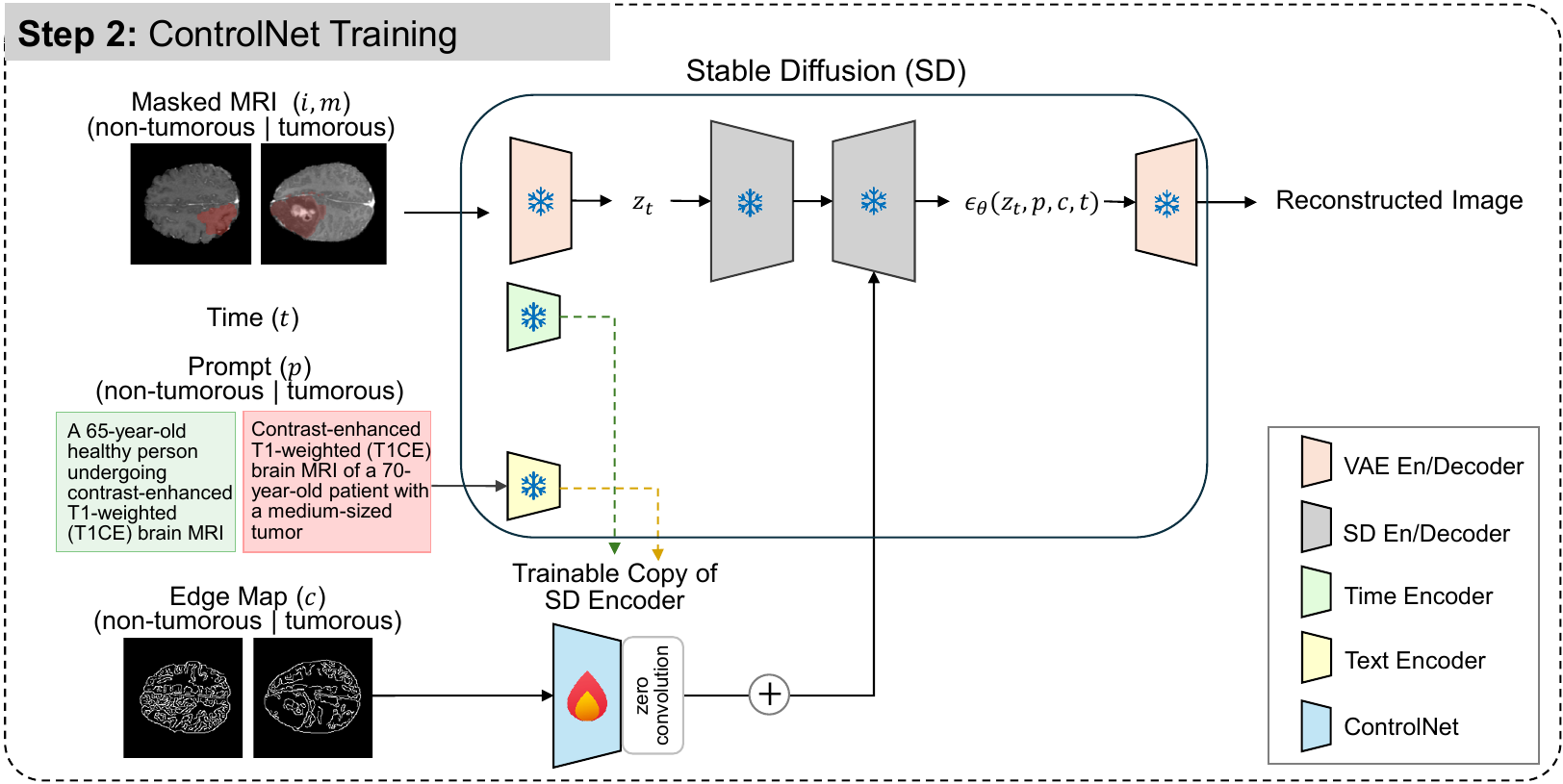}
    \caption{Step 2: ControlNet training for edge-guided anatomical reconstruction.
    ControlNet is built upon the fine-tuned SD model from Step 1.
    It receives the MRI slice, its binary mask, the corresponding prompt and diffusion time, and the edge map \(c\).
    The ControlNet is initialized with the fine-tuned SD U-Net parameters, and their learned features are transmitted to the frozen SD decoder via zero-convolution layers.
 } 
    \label{figure:step2}
\end{figure}

As shown in Figure \ref{figure:step2}, the ControlNet inherits the U-Net parameters from the fine-tuned SD model as a trainable copy, allowing it to leverage previously learned anatomical representations while processing additional edge information.
For each input pair, the corresponding edge map \(c\) is extracted from the MRI slice using the Canny edge detector~\cite{canny2009computational}. Canny edge detector is used for edge extraction due to its robust noise suppression and accurate localization enabled by Gaussian filtering and non-maximum suppression \cite{sun2022survey}.
Specifically, \(c_n\) and \(c_d\) represent the edge maps for non-tumorous and tumorous cases, respectively, paired with {\((i_n, m_n, p_n)\)} and \((i_d, m_d, p_d)\).  
These edge maps serve as conditioning inputs that capture fine anatomical boundaries relevant to each subject.  
As illustrated in Figure~\ref{figure:architecture}, ControlNet processes the edge-conditioned input to learn structural features that are injected into the frozen SD network through zero-convolution layers and residual connections. 
The zero-convolution layers are implemented as zero-initialized \(1\times1\) convolutions, allowing edge-aware features to be gradually integrated during training. 
{The injected features are scaled by a user-controlled parameter before residual addition, enabling the conditioning strength to be adjusted. This scaling allows the edge maps to serve as soft guidance.}

The overall training objective of ControlNet follows the same denoising formulation as the SD fine-tuning stage, with the addition of the edge-conditioning term:
\[
\mathcal{L}_{\text{ControlNet}} = 
\mathbb{E}_{z_t, \epsilon, t, c, m}
\left[
\|\, m \odot (\epsilon - \epsilon_\theta(z_t, p, t, c)) \,\|_2^2
\right],
\]
where \(\odot\) represents element-wise multiplication restricting the loss computation to the masked area, and \(c\) represents the edge map.
After denoising, the latent is decoded by the VAE decoder to reconstruct an anatomically coherent MRI image.

This objective allows the model to maintain the semantic consistency learned from SD while incorporating subject-specific structural cues provided by the edge map \(c\), which enables anatomically faithful and structurally guided brain reconstruction.

\FloatBarrier
\subsection{Inference: Pseudo-Healthy Brain Generation}
\label{sec:method-inference}

Pseudo-healthy brain generation is performed using the model trained through the two-stage process described in Sections \ref{sec:method-sdfinetune} and \ref{sec:method-controlnet}.
As illustrated in Figure~\ref{figure:inference}, we introduce a deliberate misalignment strategy between tumorous inputs and non-tumorous guidance to intentionally guide the model toward counterfactual pseudo-healthy brain generation.
For tumorous cases, the input configuration is deliberately adjusted to mislead the model’s reconstruction objective: the model receives the tumorous slice \(i_d\) and its binary tumor mask \(m_d\) defining the inpainting region.  
The first component of the misaligned non-tumorous guidance provides a healthy prompt \(p_h\) instantiated from the predefined template set for non-tumorous MRIs. This semantically guides the model to interpret the masked region as belonging to a non-tumorous brain. 
The second component relies on the general tendency for many major brain structures to show approximate contralateral symmetry, while recognizing that healthy brains are not perfectly symmetric and that tumors can introduce additional asymmetry \cite{kong2018mapping}.  
A mirrored edge map \(c_{\text{mirrored}}\) is generated by mirroring the contralateral hemisphere, serving as a structural reference for the missing non-tumorous region.  
By incorporating this anatomical prior, the model aligns the reconstruction with the mirrored non-tumorous boundaries, producing anatomically plausible and spatially coherent replacements for tumorous regions. {During inference, we use DDIM-based sampling with a deterministic setting to ensure consistent outputs.}
\afterpage{%
\begin{figure}[t]
    \centering
    \includegraphics[width=0.75\linewidth]{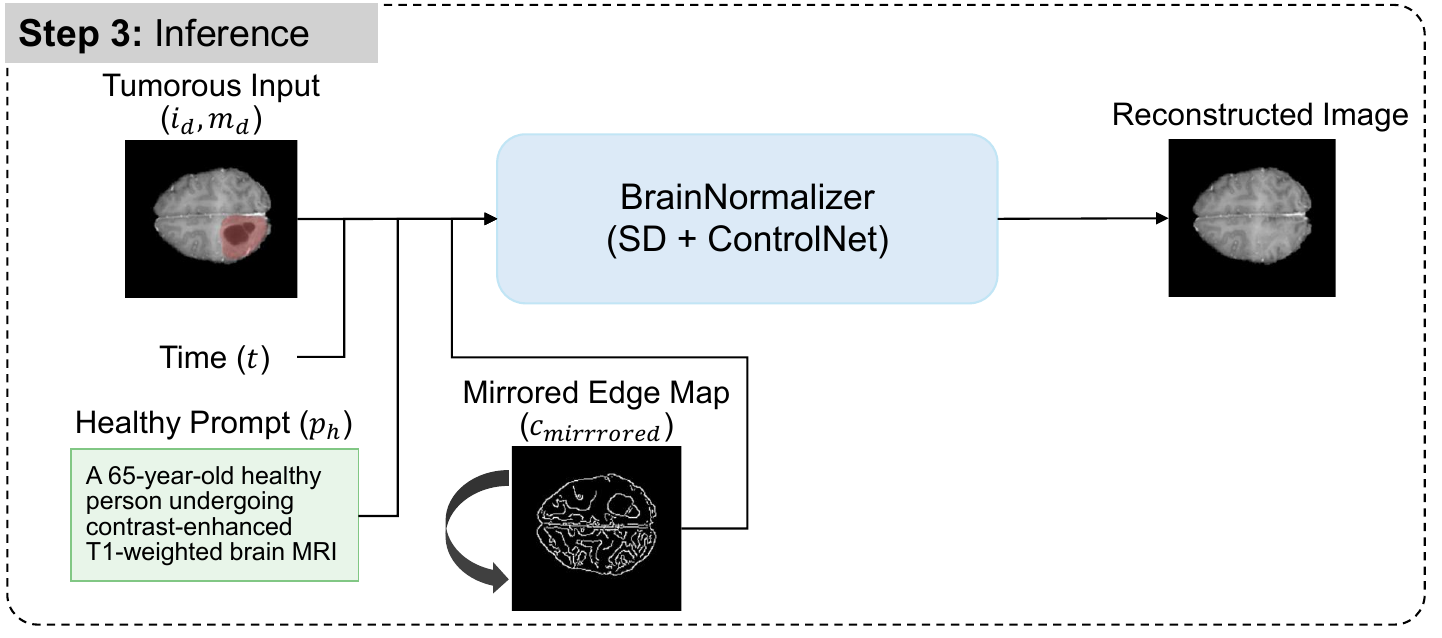}%{figures/figure_inference_1.pdf}
    \caption{
    Step 3: Inference stage of BrainNormalizer. A tumorous MRI slice with its mask is combined with a healthy prompt and a mirrored edge map to guide pseudo-healthy reconstruction.
    }
    \label{figure:inference}
\end{figure}
}

%%%%%%%%%%%%%%%%%%%%%%%%%%%%%%%%%%%%%%%%%%
\section{Experiments}
\label{sec:exp}
\subsection{Dataset and Preprocessing}
\label{sec:exp-data}
We used the BraTS2020 dataset \cite{menze2014multimodal}, which contains MRIs of brain tumor patients. {All volumes in the dataset are preprocessed with co-registration to a common anatomical space (SRI24 atlas), resulting in consistent spatial orientation and alignment across subjects \cite{baheti2021brain}. This standardization ensures that the left–right hemispheric axis is approximately aligned.} This study employed contrast-enhanced T1-weighted brain MRI (T1CE), a commonly used imaging modality in clinical practice. In the dataset, each tumor was labeled with three subregions (gadolinium-enhancing tumor, peritumoral edema, and necrotic and non-enhancing tumor core). For our analysis, these were combined into a single “tumor” region to capture the overall tumor extent. {For each 3D volume, axial slices between indices 80 and 130 were extracted, covering the central brain region where tumors are typically observed, while avoiding peripheral slices with limited or partial brain structures, following prior work \cite{wolleb2022diffusion}.} Images were clipped at the $99.5^{th}$ percentile of non-zero voxel intensities, zero-padded to a spatial size of $256\times256$, and subsequently resized to $512\times512$ using nearest-neighbor interpolation to match the SD input resolution. To align with the model’s expected three-channel input, each slice was channel-replicated. Image intensities were first normalized to the range \([0, 1]\) and subsequently rescaled to \([-1, 1]\), following the normalization procedure used in the original SD implementation.

Slices containing between 1,000 and 3,000 tumor pixels were categorized as tumorous, whereas slices with no tumor pixels were classified as non-tumorous.  
To prevent data leakage, the split was performed at the subject level, ensuring that slices from the same subject were assigned exclusively to a single set.  
The dataset was divided into training and test sets with a 9:1 ratio, maintaining the ratio between non-tumorous and tumorous slices through stratified sampling.
The training set comprised 330 subjects with 5,089 tumorous and 5,994 non-tumorous slices, resulting in a total of 11,083 slices.  
The test set included 35 subjects with 566 tumorous and 574 non-tumorous slices, for a total of 1,140 slices.
 
For edge extraction, we applied Canny edge detection after Gaussian smoothing with a kernel size of \(5 \times 5\) and a standard deviation of \(\sigma = 1.0\) to suppress high-frequency noise. The lower and upper thresholds for the Canny operator were set to 30 and 80, respectively.

\subsection{Model Configuration and Training Hyperparameters}
\label{sec:exp-model}
We adopted Stable Diffusion v1.5 (runwayml/stable-diffusion-v1-5) as the base architecture and initialized the model with publicly available pretrained weights. The model was fine-tuned in two sequential stages: (1) diffusion-based inpainting fine-tuning and (2) ControlNet fine-tuning with edge-guided structural conditioning. 

For inpainting fine-tuning, we used a batch size of 8 with gradient accumulation over four steps, resulting in an effective batch size of 32. The model was trained for 30 epochs using the AdamW optimizer ($\beta_1=0.9$, $\beta_2=0.999$) with a weight decay of 0.01. The learning rate was set to $5\times10^{-5}$, decayed to zero following a cosine schedule without warm-up steps, and the gradient norm was clipped at 1.0 to ensure training stability.  
Text prompts were tokenized using the CLIP tokenizer, and each sequence was padded to the default maximum length of 77 tokens with \texttt{[PAD]} tokens. 

For ControlNet fine-tuning, the same training configuration was adopted except for a higher learning rate of $5\times10^{-4}$ and 500 warm-up steps to facilitate faster convergence. The model was trained for 20 epochs, reflecting the reduced number of trainable parameters compared to the inpainting stage. Except for the ControlNet branch and its zero-convolution layers, all other components (VAE, CLIP text encoder, and noise scheduler) remained frozen during training. This configuration allows the model to efficiently adapt to edge-guided structural conditioning while preserving the pretrained diffusion backbone.

\subsection{Evaluation Protocol}
\label{sec:exp-metrics}
To evaluate the effectiveness of each component, we compared four models:
(1) a DDIM-based diffusion model without inpainting or ControlNet conditioning~\cite{wolleb2022diffusion},
{(2) a two-stage CycleGAN–diffusion framework for pseudo-healthy reconstruction and tumor detection~\cite{wang2024two}},
{(3) a diffusion-based inpainting model conditioned on masked MRI and tumor masks~\cite{durrer2024denoising}}, and (4) a baseline diffusion model fine-tuned only with the first-stage SD inpainting objective.
{This comparison additionally facilitates an ablation-style analysis of the proposed method. Specifically, (1) represents a diffusion model without inpainting, while (4) corresponds to an inpainting model without edge conditioning. The proposed BrainNormalizer further incorporates edge-based structural guidance, enabling a stepwise analysis of these components.}

Since paired non-tumorous ground truth is unavailable, pixel-level metrics are unsuitable for evaluating reconstruction quality, as minor pixel deviations caused by stochastic generation or noise do not necessarily indicate anatomical differences. 
To quantitatively evaluate model performance, we employed three complementary metrics that assess generative realism, downstream clinical plausibility, and structural consistency.

\textbf{Fréchet Inception Distance (FID).}
FID measures the distance between feature distributions of generated and real non-tumorous images in the feature space of a pretrained model, defined as 
$\text{FID} = \|\mu_r - \mu_g\|_2^2 + \text{Tr}(\Sigma_r + \Sigma_g - 2(\Sigma_r \Sigma_g)^{1/2})$, 
where $(\mu_r, \Sigma_r)$ and $(\mu_g, \Sigma_g)$ denote the mean and covariance of real and generated features. 
Lower values indicate closer alignment between the distributions of reconstructed and real images. {To ensure domain relevance, we compute FID using features extracted from a classifier trained on the BraTS dataset to distinguish tumorous from non-tumorous slices, with the same subject-level split as our experiments. The classifier employs a U-Net encoder architecture to obtain latent representations that capture brain MRI-specific characteristics.}

\textbf{False Positive Rate (FPR).}
{A separate U-Net-based segmentation model is used to assess the FPR, trained on the BraTS dataset with the same subject-level split.} FPR measures the proportion of reconstructed images that are predicted as containing tumors by the segmentation model and reflects the local detectability of tumor-related features. Lower values indicate that such features are less detectable, suggesting closer resemblance to tumor-free anatomy.

\textbf{Symmetry-based Structural Similarity Index Measure (SSIM).}
{Because paired non-tumorous MRIs are unavailable, we assess whether the tumor-affected region is reconstructed into anatomically plausible healthy tissue by comparing it with the contralateral non-tumorous region, which serves as a patient-specific healthy reference. This approach is motivated by the approximate hemispheric symmetry of the human brain, which provides a meaningful structural prior for assessing local anatomical plausibility. Symmetry-based SSIM evaluates structural similarity between the reconstructed region ($x$) and the contralateral healthy region ($y$), within each slice:
$\text{SSIM}(x,y)=\frac{(2\mu_x\mu_y+C_1)(2\sigma_{xy}+C_2)}{(\mu_x^2+\mu_y^2+C_1)(\sigma_x^2+\sigma_y^2+C_2)}$, 
where $\mu$ and $\sigma$ denote local means, variances, and covariance. This metric reflects symmetry-based structural consistency rather than reconstruction accuracy. Higher values indicate closer resemblance to the expected healthy structure.}

To quantify uncertainty, we repeated the evaluation using 10 random seeds and performed 1,000 patient-level bootstrap resamples over the 35 held-out test subjects for each seed. We report the mean point estimate across seeds with the 95\% bootstrap confidence interval in brackets.

\subsection{Experiment Results}
\label{sec:exp-result}

As shown in Table~\ref{table:results}, BrainNormalizer achieved the best mean performance across all metrics, with the 95\% confidence intervals indicating consistent improvements over the comparison methods. Among the comparison methods, Durrer et al. (2024) showed strong performance in terms of FID and FPR, while SD Inpaint achieved competitive symmetry-based SSIM. In contrast, Wolleb et al. (2022) and Wang et al. (2024) showed lower performance than the inpainting-based diffusion methods. This performance gap suggests that explicit conditioning on the tumor region is important for accurate pseudo-healthy reconstruction, as methods without localized tumor-region conditioning may have limited ability to selectively reconstruct pathological anatomy while preserving surrounding tissue.

{Furthermore, additional gains are observed when structural guidance is introduced. Compared to other inpainting-based methods, BrainNormalizer consistently showed improved performance, demonstrating the benefit of incorporating anatomical constraints beyond mask-based inpainting alone.}

In terms of distributional realism, BrainNormalizer achieved the lowest FID, with a mean value of 26.9, indicating the closest alignment between generated pseudo-healthy images and real healthy distributions among the comparison methods. In contrast, the CycleGAN-based method showed the highest FID of 46.8, while inpainting-based diffusion models such as Durrer et al. (2024) and SD Inpaint achieved lower FID values of 31.2 and 32.8, respectively. These results suggest that mask-based inpainting improves distributional realism, while the additional mirrored edge-guided structural conditioning in BrainNormalizer further reduces the distribution gap.

For clinical plausibility, BrainNormalizer achieved the lowest FPR of 7.1\%, indicating that its reconstructions were least likely to be identified as tumorous by the segmentation model. Compared with Durrer et al. (2024) and SD Inpaint, which achieved FPR values of 9.9\% and 10.6\%, respectively, BrainNormalizer more effectively reduced residual tumor-like detectability.

\begin{figure}[!htbp]
    \centering
    \includegraphics[width=0.95\linewidth]{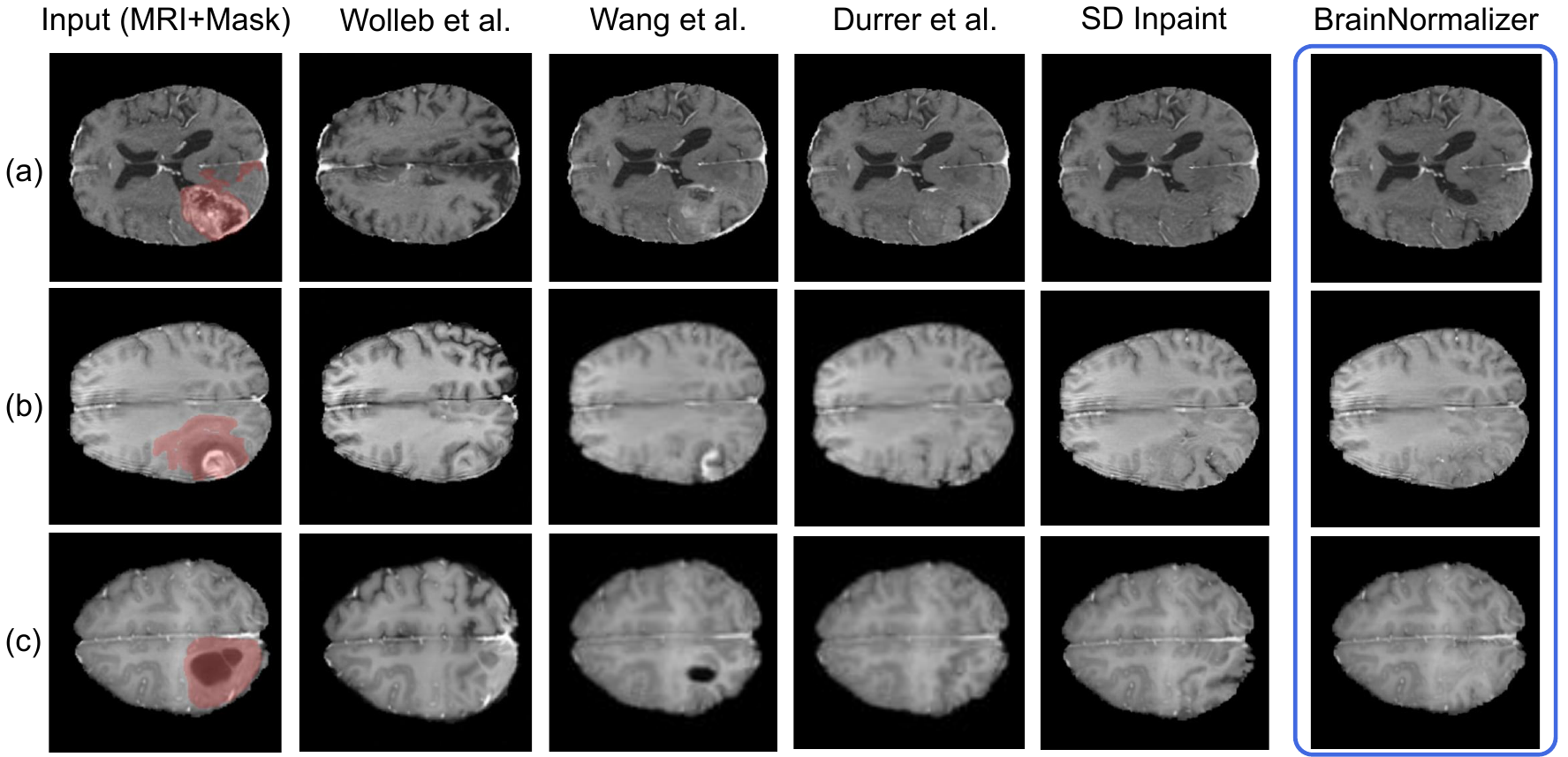}%{figures/figure2.pdf}
    \caption{Visualization of pseudo-healthy brain reconstruction results for three representative subjects. Each row corresponds to one subject ((a)–(c)). Columns show the input tumorous MRI with tumor mask overlay and the reconstructed images produced by each method.} 
    \label{figure:visualization}
\end{figure}

BrainNormalizer also achieved the highest symmetry-based SSIM of 0.74, indicating stronger structural consistency with the contralateral healthy reference. This result indicates that mirrored edge guidance improves anatomical consistency beyond mask-based inpainting alone, which is also supported by the qualitative results in Figure~\ref{figure:visualization}.
\begin{table}[!htbp]
\centering
\caption{Quantitative comparison of generative performance across different models. Values are reported as mean [95\% CI]. For each metric, lower FID and FPR values and higher SSIM* indicate better performance. *SSIM is a symmetry-based metric computed against the contralateral healthy region, as ground-truth pseudo-healthy images are not available.}
\label{table:results}
\resizebox{\textwidth}{!}{
\begin{tabular}{lccc}
\hline
\textbf{Model} 
& \textbf{FID} ($\downarrow$)
& \textbf{FPR (\%)} ($\downarrow$)
& \textbf{SSIM*} ($\uparrow$) \\
\hline
Wolleb et al. 
& 42.9 [42.1, 43.7] 
& 15.8 [15.0, 16.7] 
& 0.37 [0.36, 0.39] \\

Wang et al. 
& 46.8 [45.5, 48.0] 
& 18.0 [16.8, 19.0] 
& 0.28 [0.26, 0.31] \\

Durrer et al.
& 31.2 [30.5, 32.0] 
& 9.9 [9.5, 10.6] 
& 0.59 [0.56, 0.61] \\

SD Inpaint 
& 32.8 [32.4, 33.4] 
& 10.6 [10.1, 11.0] 
& 0.67 [0.66, 0.69] \\

\textbf{BrainNormalizer (Proposed)} 
& \textbf{26.9 [26.5, 27.3]} 
& \textbf{7.1 [6.7, 7.4]} 
& \textbf{0.74 [0.72, 0.75]} \\
\hline
\end{tabular}
}
\begin{flushleft}
\end{flushleft}
\end{table}

Figure~\ref{figure:visualization} presents pseudo-healthy reconstruction results for three representative subjects. BrainNormalizer produces more anatomically consistent reconstructions while effectively removing tumor regions.
The DDIM-based model (Wolleb et al. (2022)) often left residual tumor signals and altered unrelated regions, {and the CycleGAN-based approach (Wang et al. (2024)) further exhibited structural inconsistencies and unrealistic artifacts.
In contrast, inpainting-based diffusion methods (SD Inpaint and Durrer et al. (2024)) generated more localized modifications within tumor regions. However, they still show incomplete tumor removal (Figure~\ref{figure:visualization}(b),(c)) and noticeable asymmetry in structurally deformed regions (Figure~\ref{figure:visualization}(a)).}
By incorporating edge-guided conditioning, BrainNormalizer achieves more symmetric and anatomically faithful reconstruction, with improved structural consistency (Figure~\ref{figure:visualization}(a)) and more complete tumor removal (Figure~\ref{figure:visualization}(b),(c)).

\begin{figure}[!htbp]
    \centering
    \includegraphics[width=0.70\linewidth]{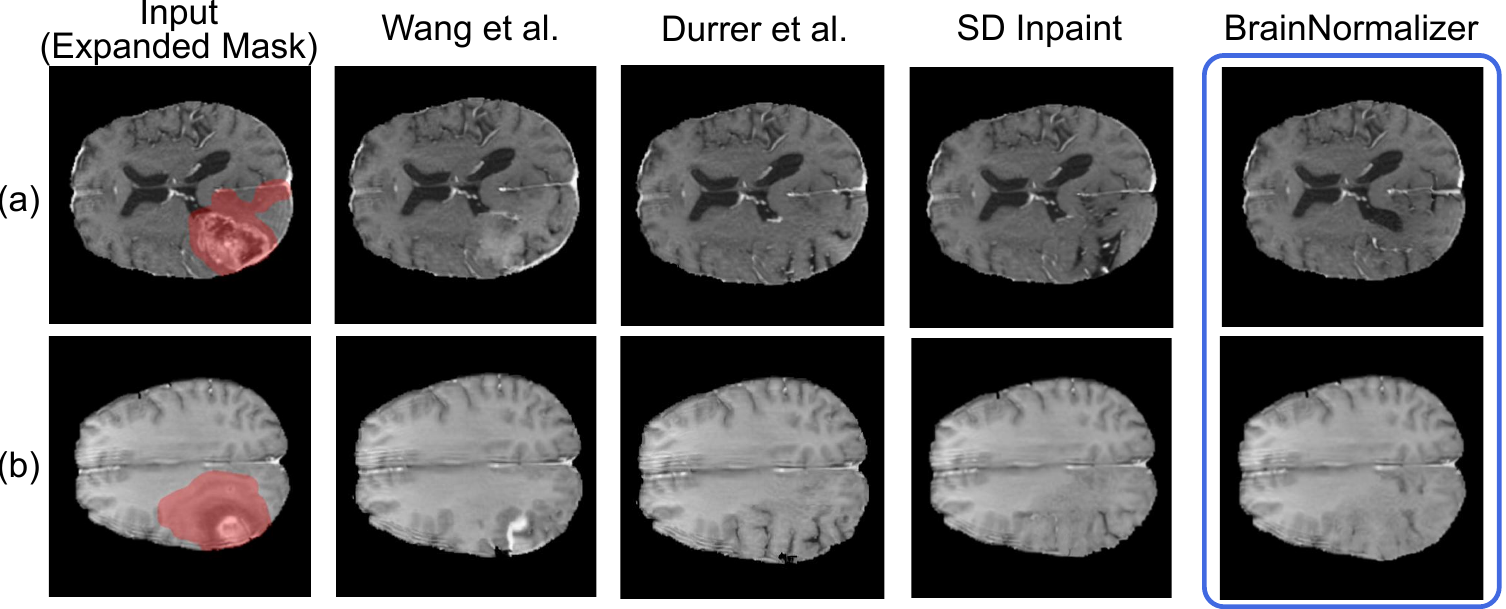}
    \caption{Visualization of reconstruction robustness to mask size. The same input MRIs as those in Figure~\ref{figure:visualization}(a)–(b) are used, but with expanded tumor masks. Each row corresponds to one subject. Columns show the input MRI with expanded mask and the reconstructed results from each method.} 
    \label{figure:mask}
\end{figure}

Inpainting-based methods require tumor masks, which may vary in accuracy in practice. To evaluate robustness to mask quality, we performed pseudo-healthy reconstruction using expanded tumor masks to simulate coarse annotations, as shown in Figure~\ref{figure:mask}. The same subjects as in Figure~\ref{figure:visualization}(a) and (b) are used for direct comparison.
BrainNormalizer produces consistent reconstructions with those obtained using default masks, preserving symmetry and effectively removing tumor regions.
{In contrast, other methods exhibit degraded performance under expanded masks, including incomplete tumor removal, reduced symmetry (Figure~\ref{figure:mask}(a)), and visible artifacts such as structural discontinuities (Figure~\ref{figure:mask}(b)).
These results suggest that BrainNormalizer is more robust to mask imprecision, indicating its potential to reduce the burden of precise tumor annotation in practical settings.}
\FloatBarrier

Table~\ref{table:tumor_size} summarizes BrainNormalizer performance across tumor-size groups. BrainNormalizer maintained relatively stable performance across tumor-size groups, with overlapping or comparable 95\% confidence intervals across FID, FPR, and symmetry-based SSIM. These results suggest that reconstruction quality did not show a clear monotonic degradation with increasing tumor size in the evaluated test set.
\begin{table}[!htbp]
\centering
\small
\caption{BrainNormalizer performance across different tumor sizes. Values are reported as mean [95\% CI]. Tumor size categories are defined based on tumor pixel counts. Lower FID and FPR values and higher SSIM* indicate better performance. SSIM* denotes symmetry-based SSIM computed against the contralateral healthy region, as ground-truth pseudo-healthy images are not available.}
\label{table:tumor_size}
\setlength{\tabcolsep}{4pt}
\renewcommand{\arraystretch}{1}
\begin{tabular}{lccc}
\hline
\textbf{Tumor size} 
& \textbf{FID} ($\downarrow$)
& \textbf{FPR (\%)} ($\downarrow$)
& \textbf{SSIM*} ($\uparrow$) \\
\hline
Small ($<$ 1,400) 
& 27.6 [27.1, 28.0] 
& 7.6 [7.2, 7.9] 
& 0.77 [0.75, 0.78] \\

Mild (1,400--1,800) 
& 27.2 [26.7, 27.7] 
& 7.7 [7.0, 8.1] 
& 0.74 [0.72, 0.76] \\

Medium (1,800--2,100) 
& 28.2 [27.8, 28.6] 
& 7.4 [6.7, 7.8] 
& 0.73 [0.71, 0.74] \\

Moderate (2,100--2,500) 
& 27.1 [26.7, 27.5] 
& 8.0 [7.7, 8.2] 
& 0.75 [0.73, 0.76] \\

Large ($>$ 2,500) 
& 26.0 [25.6, 26.8] 
& 6.3 [5.7, 6.7] 
& 0.78 [0.76, 0.79] \\
\hline
\end{tabular}
\end{table}

\subsection{Ablation Study}
\label{sec:exp-ablation}

To further examine the contribution of each component in BrainNormalizer, we conducted a component-wise ablation study by varying the model components and guidance signals. The ablation was designed to disentangle the contributions of four design choices: the role of mask-based inpainting in localizing reconstruction to the tumor region, the role of ControlNet-based edge conditioning in enforcing anatomical structure, the effect of using original versus mirrored edge maps as structural guidance, and the importance of healthy semantic prompting for pseudo-healthy generation. All variants were evaluated on the same held-out test set using the same tumor masks and evaluation metrics.

As shown in Table~\ref{table:ablation}, SD alone showed limited structural consistency, with a low symmetry-based SSIM of 0.24. Adding mask-based inpainting improved SSIM to 0.67 and reduced FPR from 17.2\% to 10.6\%, indicating that localized reconstruction is important for preserving surrounding subject-specific anatomy. However, ControlNet with the original edge map degraded performance, suggesting that tumor-induced boundaries in the original edge condition can misguide pseudo-healthy reconstruction. Moreover, using the mirrored edge map with a tumor prompt resulted in the highest FPR and the lowest SSIM, showing that the healthy prompt plays an essential role that the mirrored edge map alone cannot compensate for. The full BrainNormalizer configuration achieved the best overall performance, with an FID of 26.9, an FPR of 7.1\%, and a symmetry-based SSIM of 0.74. These results support the complementary roles of mask-based inpainting, mirrored anatomical guidance, and healthy semantic prompting in pseudo-healthy reconstruction. Confidence intervals for all metrics are reported in Table~\ref{table:ablation}.
\begin{table}[!htbp]
\centering
\caption{Component-wise ablation study of BrainNormalizer. Values are reported as mean [95\% CI]. Lower FID and FPR and higher SSIM indicate better performance. SSIM* denotes symmetry-based SSIM computed against the contralateral healthy region.}
\label{table:ablation}
\setlength{\tabcolsep}{0.95pt}
\renewcommand{\arraystretch}{0.80}
\fontsize{6.5}{9}\selectfont

\begin{tabular}{lccccccc}
\hline
\textbf{Model} 
& \textbf{Inpaint} 
& \textbf{ControlNet} 
& \textbf{Edge} 
& \textbf{Prompt} 
& \textbf{FID} ($\downarrow$)
& \textbf{FPR (\%)} ($\downarrow$) 
& \textbf{SSIM*} ($\uparrow$) \\
\hline

SD 
& -- & -- & None & Healthy 
& 38.4 [37.6, 39.4] 
& 17.2 [16.3, 18.1] 
& 0.24 [0.22, 0.26] \\

SD Inpaint 
& \checkmark & -- & None & Healthy 
& 32.8 [32.4, 33.4] 
& 10.6 [10.1, 11.0] 
& 0.67 [0.66, 0.69] \\

ControlNet (Original Edge)
& \checkmark & \checkmark & Original & Healthy 
& 43.7 [42.8, 44.8] 
& 16.5 [15.6, 17.5] 
& 0.32 [0.30, 0.35] \\

ControlNet (Tumor Prompt)
& \checkmark & \checkmark & Mirrored & Tumor 
& 45.5 [44.3, 46.7] 
& 46.7 [45.2, 48.3] 
& 0.18 [0.16, 0.20] \\

\textbf{BrainNormalizer}
& \checkmark & \checkmark & Mirrored & Healthy 
& \textbf{26.9 [26.5, 27.3]} 
& \textbf{7.1 [6.7, 7.4]} 
& \textbf{0.74 [0.72, 0.75]} \\
\hline
\end{tabular}
\end{table}

%%%%%%%%%%%%%%%%%%%%%%%%%%%%%%%%%%%%%%%%%%
\FloatBarrier
\section{Conclusions}
\label{sec:conclusion}
We presented BrainNormalizer, a diffusion-based framework for generating anatomically plausible pseudo-healthy brain MRIs from tumorous scans without requiring paired data. The proposed approach integrates edge-guided ControlNet conditioning to enforce structural consistency and introduces a deliberate misalignment strategy that enables subject-specific, anatomy-informed reconstruction at inference. By leveraging patient-specific anatomical cues, BrainNormalizer produces pseudo-healthy references that can support tumor delineation and downstream analysis. Experiments on BraTS2020 demonstrate improved perceptual realism, structural fidelity, and clinical plausibility over existing methods, with qualitative results indicating consistent reconstruction of subject-specific anatomical structures.

While demonstrating promising results, BrainNormalizer exhibits certain limitations. {Reconstruction quality degrades when tumors occupy large central regions or affect both hemispheres simultaneously.
It also degrades when anatomical asymmetry is pronounced.} {In these cases, the mirrored edge map may become less reliable, and strong symmetry guidance can introduce structural artifacts or suppress natural anatomical asymmetry. These failure modes are further discussed in Appendix B. To mitigate this, a user-defined conditioning scale allows balancing structural guidance with the learned generative prior.} {Furthermore, variations across datasets, such as differences in imaging protocols and patient populations, may introduce distribution shifts.}

Future work will explore several directions to address these challenges. First, integrating volumetric context through 3D diffusion backbones could improve inter-slice consistency and reduce limitations of the current 2D setting. Second, incorporating adaptive cross-hemisphere attention mechanisms may enable the model to better handle asymmetric tumor presentations. Third, extending the framework to multi-contrast MRI modalities and improving robustness to distribution shifts are important directions for future work. Finally, clinical validation studies with radiologists and neurosurgeons would be essential to assess the practical utility of generated pseudo-healthy brain references in real-world treatment planning and surgical guidance workflows.

\section*{Disclosure statement}
The authors report there are no conflicting interests to declare.

\section*{Funding}
%Funding information is withheld for double-blind review.
We are grateful to the National Cancer Institute for supporting this work (U01CA250481, U54CA274504).

%\section{References}
\bibliographystyle{unsrtnat}
\bibliography{references}

%\texttt{thebibliography}

\FloatBarrier
\appendix
\renewcommand{\thesection}{Appendix~\Alph{section}}
\section{Sensitivity Analysis}
\label{sec:sen}
{We analyze the sensitivity of BrainNormalizer to key hyperparameters in the diffusion and ControlNet components. The default configuration used in our experiments is as follows: number of diffusion steps = 50, noise strength ($s$) = 1.0, classifier-free guidance scale ($g$) = 7.5, and ControlNet conditioning scale ($c$) = 0.8.}

\begin{figure}[!htbp]
    \centering
    \includegraphics[width=0.90\linewidth]{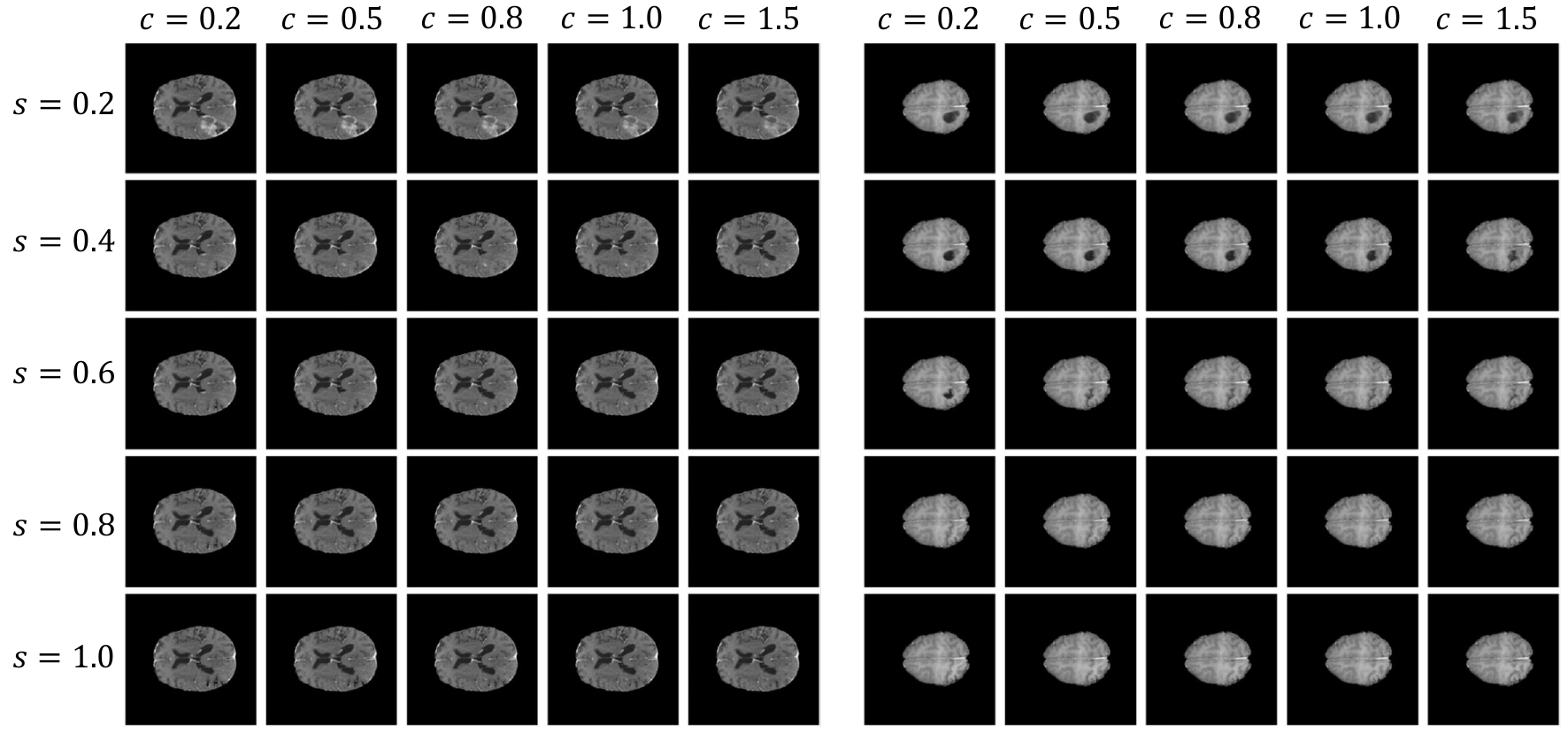}
    \caption{Qualitative sensitivity analysis for diffusion strength ($s$) and conditioning scale ($c$).} 
    \label{figure:sa}
\end{figure}

{Figure~\ref{figure:sa} shows qualitative results under different combinations of $s$ and $c$, with other parameters fixed to the default settings. Across the expanded range, BrainNormalizer demonstrates stable reconstruction behavior, with consistent anatomical structures preserved across most parameter settings. The denoising strength $s$ primarily influences the overall intensity and smoothness of the reconstructed regions, while the conditioning scale $c$ governs the degree of structural adherence to the edge guidance. Moderate values of $c$ (e.g., around 0.8–1.0) consistently yield improved structural consistency across different $s$ levels, while further increasing $c$ leads to diminishing returns without noticeable additional improvement. Similarly, very low $s$ values result in insufficient reconstruction, whereas higher values maintain stable outputs without significant degradation.}
\FloatBarrier

\section{Failure Case Analysis}
\label{sec:failure}
\begin{figure}[!htbp]
    \centering
    \includegraphics[width=0.90\linewidth]{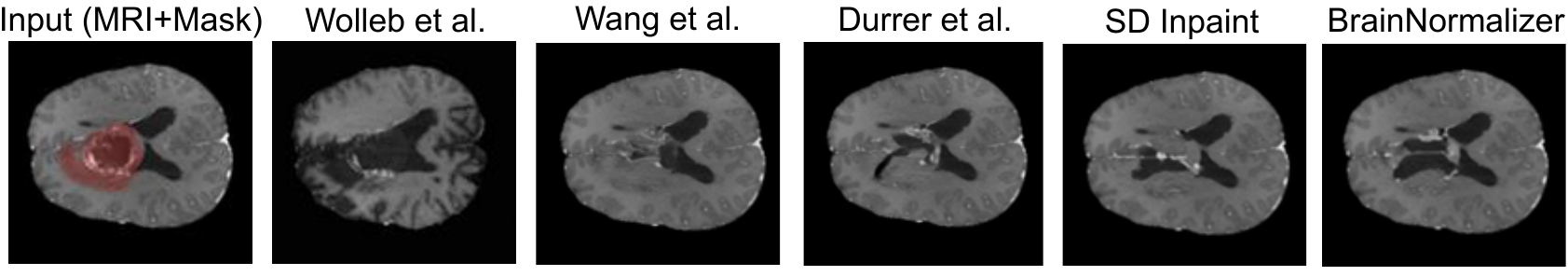}
    \caption{Failure case on MRI with midline tumor: pseudo-healthy reconstruction results across methods.} 
    \label{figure:failure}
\end{figure}

{As illustrated in Figure \ref{figure:failure}, in the challenging midline tumor case, where the mirrored edge map does not provide accurate healthy structural guidance, BrainNormalizer exhibits degraded reconstruction quality. Nevertheless, it better preserves the overall morphology of the lateral ventricles and mitigates tumor-affected regions compared to other methods. In contrast, other methods show more pronounced structural distortions and less coherent reconstructions.}
\FloatBarrier

Beyond midline tumors, comparable limitations may arise in cases with bilateral tumor involvement, pronounced natural asymmetry, or severe mass effect, as the mirrored contralateral edge map may no longer provide a reliable healthy anatomical reference.
This concern is supported by the component-wise ablation study in Section \ref{sec:exp-ablation}, which showed that reconstruction quality depends on both the edge map used for ControlNet conditioning and the semantic prompt. In bilateral cases, both hemispheres may contain pathological structures, making it difficult to use the contralateral side as a pseudo-healthy reference. In cases with strong natural or tumor-induced asymmetry, mirrored guidance may over-constrain the reconstruction and suppress subject-specific anatomical differences. In cases with large mass effect, displacement of midline or contralateral structures can reduce the validity of the mirrored-edge assumption and lead to less reliable pseudo-healthy reconstruction.
\end{document}